\documentclass{article}

\title{On the theory of system administration}
\author{Mark Burgess}
\date{22 March 2000}

\usepackage{psfig}

\def\boxx{\vcenter{\vbox{\hrule height.4pt
          \hbox{\vrule width.4pt height8pt
          \kern8pt\vrule width.4pt}\hrule height.4pt}}}

\def\beq{\begin{eqnarray}}
\def\eeq{\end{eqnarray}}
\def\2{\frac{1}{2}}
\newtheorem{axiom}{Basic assumption}
\newtheorem{theorem}{Theorem}

\begin{document}
\maketitle

\begin{abstract}
This paper describes necessary elements for constructing theoretical
models of network and system administration. Armed with a theoretical
model it becomes possible to determine best practices and optimal
strategies in a way which objectively relates policies and assumptions
to results obtained.  It is concluded that a mixture of automation and
human, or other intelligent incursion is required to fully implement
system policy with current technology. Some aspects of the author's
immunity model for automated system administration are explained, as
an example. A theoretical framework makes the prediction that the
optimal balance between resource availability and garbage collection
strategies is encompassed by the immunity model.
\end{abstract}

\section{Introduction}

System administration is the realm of computer science which deals with
the planning, configuration and maintenance of computer systems. It is
presently a discipline founded mainly on the anecdotal experiences of
system managers\cite{evard1}.  To date, no formal (mathematical) analyses of
system administration have been undertaken, with the aim of making
more scientific studies. This makes it difficult to express objective
truths about the field, avoiding marketing assertions and the vested
interests of companies and individuals, which are common in the
commercial sector.

The aim of the present work is to establish a formal basis for the
field, a way of formulating a framework for objective discussions
about computer management. It will hopefully serve as a bridge between
mathematical disciplines and system administrators. In this respect,
the paper may be viewed mainly as a commentary, laying some
foundations for future work, rather than providing immediate solutions.

In previous work, it has been shown how the average behaviour of
systems of computers and users can be approximated by a blend of
statistical models and thermodynamical ideas\cite{burgessLISA99}. That
work allows us to form a mathematical model of computer systems which
can be used as a basis for modelling system administration.  The study
of computer behaviour has much in common with the physics of
thermodynamics. From a coarse mathematical viewpoint, system
administration can be viewed in much the same way as thermodynamical
experiments with pistons and engines, i.e. moving information and
resources around in such a way as to change the state of the
system. However this viewpoint is mainly useful in a calculational
setting. System administration also has much in common with
medicine. In many ways, system administration is medical science for
computers: a somewhat simpler problem than that of human physiology,
but nonetheless involving many of the same themes: nutrition,
regulation, immunity and repair.

What then should a theory of system administration be about?  The task
of elucidating this sounds straightforward, but it is a slippery
business. System administration, in reality, is based on mainly
qualitative, high level concepts, which mix technical and sociological
issues at many levels. Although it is clear to system administrators
that there is a body of technical principles involved in the
discipline, it remains somewhat intangible from the viewpoint of a
scientist.  It is hard to find anything of general, reproducible value
on which to base a more quantitative theory.

One of the obstacles to formulating such a theory is the complexity of
interaction between humans and computers.  There are many variables in
a computer system, which are controlled at distributed
locations. Computer systems are {\em complex} in the sense of having
many embedded causal relationships and controlling
parameters. Computer behaviour is strongly affected by human
social behaviour, which is often unpredictable.  The task of
identifying and completely specifying the ideal state is therefore a
non-trivial one. It is nonetheless this task which this paper attempts
to address. Can one formulate a quantitative theory of system administration,
which is general enough to be widely applicable, but which is specific
enough to admit analysis?

If this, already significant problem can be addressed in sufficient
terms, one might then aim to look further towards general regulatory
systems and approach more ambitious questions. It is not difficult to
see many analogous questions in other areas of science, which could be
applicable to system administration. For instance: what is the
effectiveness of generalized immunity and repair
systems\cite{burgessLISA98} (automatic repair and regulation)? Is
there an optimal strategy for error detection and correction? Is a
system administrator's human mind (playing the role of doctor/surgeon)
better or worse than a mechanistic response or immune system?  This
last point is often a bone of contention in the system administration
community. Should tasks be automated? Or should a human lawgiver
always remain in manual control? What is more efficient?  Biological
systems point to the need for both types of management: at any given
moment, a doctor's intelligence and superior human cognition can
compensate for a lack of adaptation in our programmed immune
responses, but the automatic immune response is both faster and more
capable than a doctor when its program is sufficient.  Certainly the
empirical evidence in biological information systems is compelling:
after billions of years of evolution, nature has established immune
systems in all vertebrates larger than a tadpole. Of course, this is
no indication that the solution is optimal. No acceptable analysis has
been used to demonstrate this yet.  It could be that vertebrate
evolution is merely poised on some plateau between minima of much
deeper importance.

The aim of this paper then is to elevate system administration from an
expression of subjective opinion to a more objective, scientific
level, hopefully without inflating it meaninglessly into
pseudo-science or philosophy.  In order to limit the length of this
paper, solutions of the models and constraints will be kept to a
minimum here. However, it will be possible to draw a few general
conclusions, even without reference to specific models.

The outline of this paper is as follows. To begin the discussion it is
necessary to establish some basic axioms.  It is important to restrict
the scope of what a theory of system administration encapsulates;
without such a restriction, one ends up with either many disjointed
pieces or only vague hand-waving notions.  Having determined the ground
rules, it is then appropriate to identify the basic operations which
can be carried out within that scope. This identification is required
in order to formulate a discussion of strategies for system
management.  Once this level of formality has been attained,
strategies can be formulated, based on types of action and timing and
the task of administrating a computer system can be described in
precise game theoretical terms. This is the primary goal of this work.

\section{The scope of system administration}

One of the first obstacles in discussing the theory of system
administration is defining its scope. System administrators are called
upon to perform all manner of tasks as part of their duties. This
battery of skills has no particular cohesion or structure to it, so it
resists formalization. We must improve on this situation if we are to
make progress in forming a theory of system administration.  In
particular, we must restrict its scope to encompass only core
activities. These core activities will include insuring availability,
efficiency, and security for all users, and finally fault diagnosis of
the system. This includes issues such as software installation
and upgrades, which can be classified under availability and efficiency.
It also includes user management to a certain extent, though it will not
be useful to address the issue of creation of user accounts in this
context.

\section{On scales}

A well known feature of descriptions of complex systems is that a
complete understanding is best organized as a unification of the
partial understanding of the system at several different levels or
scales. Complex systems are often so disparate at different scales
that quite different descriptions are required to capture the full
essence.  A theory of system behaviour at, say the microscopic level
of system calls, need not resemble a theory for the behaviour at a
macroscopic scale of larger entities, such as patterns of user
behaviour. Both are needed in order to understand the whole hierarchy
of things going on.

If one is only interested in high level phenomena, then the details of
low level phenomena are seldom directly relevant, to a good
approximation. This is the principle of separation of scales. The
principle states that, as one moves from microscopic to macroscopic
scales, new behaviour can emerge as collective phenomena, which often
depends only weakly on the microscopic details of the levels below.
This is a simple idea, which is quite intuitive, but which has far
reaching consequences. It can easily be appreciated with the help of a
couple of examples.

A bridge, for instance, has the property of spanning a distance and
carrying weight, regardless of whether it is made from steel or copper
or wood. The choice of material and the microscopic arrangement of
atoms in the metal or wood, of course, tells us something about the
strength of the bridge, but perhaps not as much as the structure of
the bridge at the scale of the whole thing. In other words, the construction
of the bridge at the scale of the users of the bridge is far more important
to its function than the microscopic construction of its pieces under a microscope.

Similarly, to an acceptable approximation, the behaviour and operation
of a sales database, at the level of information transactions (their
order and type), is more important to an information retrieval system
than how those transactions are implemented through system calls
(e.g. whether the system runs on Unix or on NT). The ability to
retrieve information does not depend on whether the storage medium is,
an IDE or a SCSI disk.  The same job will be done regardless.

To summarize, a description of system behaviour at a high level is,
for many purposes, independent of specific details of the lower
levels.  Computer systems can be modelled by generic computer systems
with certain high level characteristics; similarly users can be
modelled as idealized users, also with common characteristics. A theory of
system administration will be most successful if it appeals to such
generalities, rather than delving into unnecessary specifics.

\section{Axioms of system administration}

To begin a formal discussion, we need to establish a frame of
reference, i.e. the ground rules for the discussion. In this section,
a basic fundament is proposed with the aim of striking a balance between
reality and suitability for analysis. It is also necessary to
partially limit the scope of the discussion to avoid unnecessary
complication. Although the aim of this presentation is not precise
mathematical rigor, it is the aim to indicate that such a rigor is
possible and to indicate how it can be provided. A secondary aim is to
communicate the key elements of the discussion to a more theoretical
audience; for these reasons, the language adopted is one which is
meant to build bridges between system administration and more
mathematical disciplines. Readers are asked to keep an open mind with
regard to use of terms however, since technical disciplines often use
words in meanings which are specific to those disciplines, and this could
lead to confusion.

A computer system is analogous to a community\cite{burgessbook}
composed of many interacting and competing players: i.e. users and
administrators. It can only properly be discussed in terms of the aims and
activities of this collective and of individual members of that
community. Not all the members of a community share the same
objectives, as a general rule.  Traugott and Huddleston have pointed
out\cite{NASAsteve} that it is often pertinent to view a local
computer community as a single virtual machine, rather than as a
conglomeration of individual hosts. In this paper, the term computer
system will be used to refer to the collective hosts of a local
domain, or some appropriate logical unit of networked computers. It is
taken for granted that there may be internal competition for resources
and even conflict between competing parties.

In order to formulate a theory of system administration we must
establish a set of possible goals, procedures and obstructions and
state them in formal terms. The aims and intentions of each computer
system are different; usually they are prescribed by a system policy,
i.e. a formal statement of intent and allowed practice.  The aim is
then to postulate or derive strategies which best achieve those goals,
given the essential constraints. From this viewpoint, one expects
the language of constrained competition to play a role in a theory of
system administration.  Even if one could frame such a theory in
formal terms, what would be the purpose of such an exercise? The
principal benefit of such an attempt is to create a rigid protocol for
discussing system administration, which is general enough to cover most
of the actual problems and possibilities, but which is stringent
enough to prevent its perversion by parties with vested interests in
proving a certain point of view. 

There is a number of stages in this programme. To begin with, one
needs some basic axioms which all parties agree on, propositions which
define the aims of system administration. Next one needs to abstract a
model of a computer system which is sufficient to capture the
dynamical interaction between all of the players, but which is
sufficiently simple to be surrendered for analysis.  Here we shall
suppose that a computer's resources (memory, CPU, disk etc) are
divided into two parts, 
\beq
R = R_c \oplus R_m,
\eeq
i.e. a part which determines the behavioural configuration parameters
$C$ of the system working resources, and a remainder part (the working
resources themselves) which users of the system can change as a normal
part of their interaction with the system. This remainder part can be
observed over an appropriate time scale, giving a set of measurements
$\overline M$ which indicate how the system is being used.

The different possible configurations of the system resources $\lbrace C
\rbrace$ are made up from the independent operation types $\lbrace T \rbrace$
which lead to these configurations.  From this definition one needs to
be sure that a unique description is possible, i.e. eliminate points
of contention about the description itself.  Finally, one must be sure
that the description is sufficiently complete, i.e. that there exists
a mapping between policy and system configuration which is as complete
as the problem itself. The purpose of this section is to introduce the
key players in this description, in advance of a fuller description in
the coming sections.

In order to state the purpose of system administration, we may take
the basic tenet or principle to be the following:
\begin{axiom}
{\it The requirements and constraints of any computer system are
defined at any time $t$ by an implementable
system policy $P(t)$. This policy determines the actions or rules of
play for a system administrator, but not necessarily the actions of
users. It includes a specification of which and how many users are
allowed to access the system.}
\end{axiom}
The policy $P(t)$ is not usually a continuous function of time, but
may change catastrophically (in the mathematical sense) over a time
scale which is much longer than the time scale over which users act and
make changes to the system. The nature of this policy is not yet determined.

In order to make a policy implementable, it must be possible to relate
it to a {\em complete configuration instruction} for the system
$C(t)$, through rules and constraints. These rules and constraints
could be issued verbally to users, or could be programmed into
configuration files of software components which form the system.  A
single complete configuration instruction for the system can be thought of as
being a sum of two parts:
\beq
C = C_r \oplus C_u,
\eeq
a specification of resource configurations $C_r$ which describe how
the software and hardware landscape is configured, and a specification
of user configurations $C_u$, which describes who is allowed to do
what with the resources (this includes remote, network users who access
services through local agents). A specification of user configurations $C_u$
(numbers of users and their rights to resources) could easily be
separated from system policy conceptually, but it is convenient to
view the policy as a complete specification of the system plus its
intended and actual usage.  The meaning of the symbol $\oplus$
is that of a heuristic union: configuration specifications take
many forms (are objects of many types).  They are most easily thought
of as sets of more primitive objects, in which case the addition of
sets implies their strict union.

A {\em complete configuration instruction} can be thought of
geometrically as a point in a vector space, which is found by adding
together instructions of linearly independent (orthogonal) types. One
does this by introducing a set of {\em primitive configuration
instruction types} $\lbrace T^i \rbrace$, and writing the complete
configuration as a linear combination of these:
\beq
C = \sum_i \; c_i T^i,
\eeq
with set-valued coefficients $c_i$.
The basis of primitive configuration operations will be described
later. A complete configuration usually contains instructions for
the operators of the system also. The system administrator can also be
viewed as part of this system for the sake of abstraction.

The set of all configurations $\lbrace C \rbrace$ contains
much redundancy. Let us imagine that the mapping of complete
configuration instructions to the final state of the system is many to
one, and that this multiplicity can be represented by a group of
permutations and transformations ${\cal G}$\footnote{The nature of
this group could be fairly complicated and is not particularly
important to the discussion. The fact that the redundancy, in
principle, may be represented by a mathematical group is an
idealization which is attainable in in theory. It is not an expression
of the current state of affairs in the world of computers.}. Thus
equivalent configurations could be formed by permuting configuration
instructions, if ordering is unimportant, or by exchanging
(transforming) one component of the configuration for another.  An
example of this is the following: the configuration of a World Wide
Web service might be possible with several equivalent software
systems, each with equivalent configuration files: in this case, these
would form an equivalence class.  Conversely, if even a minor detail
distinguishes them, then they are inequivalent configurations.

Let us define an implementable policy $P(t)$ as being any
representative member of the set of equivalent configurations $\lbrace
C(t)\rbrace$
\beq
P(t) \equiv \lbrace C(t) \rbrace / {\cal G}.
\eeq
A policy could naturally mean more than a configuration (or computer
and its operators), but as long as other aspects of the policy cannot
be implemented by either machine or human, they are irrelevant to the
system.  Having related policy to configuration instructions, the path
is clear to define the state $S(t)$ of the system.

Let a state of the system describe a single configuration of users $C_u$, of
system resources $C_r$ and a set of measured average metrics
$\overline M$ which summarizes the average usage of the system in
relation to the users\cite{burgessLISA99}. The metrics $\overline M$ represent a first
order response (feedback interaction) between users and resources.  The
state is written, again, as a direct sum
\beq
\overline S_p \equiv \left(\frac{C_r \oplus C_u}{\cal G}\right) \oplus \overline M(C_r,C_u)
\eeq
where the behavioural metrics $\overline M(C_r,C_u)$  
are functions of resources and user activity. One may now state the following
provable hypothesis.

\begin{theorem}
{\it Any sufficiently complete system policy $P(t)$ specifies, by
implication, a representative average ideal state $\overline S_p(t)$,
from an equivalence class of ideal states $\lbrace\overline S\rbrace$ under $\cal G$,
for the computer system concerned, over user time-scales $T_u$, provided
that the rate of change of policy $dP/dt$ is much smaller than changes
in user behaviour $d\overline M/dt$, i.e. the policy changes on the order of weeks or months
rather than hours or days.

\noindent {\rm\bf Corollary}:  The ideal state can only be identified on
average, since interactions with unpredictable user activity are
constantly causing fluctuations $\delta S(t)$ in the state of the
system.  These fluctuations also occur at a rate $d\overline M/dt \gg dP/dt$.}
\end{theorem}

The existence of an ideal state has already been used in designing the
author's site Configuration Engine (cfengine)\cite{burgess1}, but it
has not previously been explained at length.  The proof of this
theorem is straightforward, from the definitions. Every computer
system has a finite set of resources and configuration objects which
is completely prescribed by a total configuration $C_p$. Each resource
object may be in a state described by a finite length bit string,
describing a distinct configuration $s_i \in C_r$. There is therefore
a mapping from the configuration $C_p$ to the actual state
\beq
C_p \rightarrow \overline S_p + \delta S.
\eeq
This mapping is one to many, since $\delta S$ is a stochastic
variable.  The averaging operation eliminates the non-uniqueness by
extinguishing $\delta S$, provided the averaging process is defined,
i.e., $d\delta S/dt \gg dP/dt$.  Thus any complete set of average
measurements contributes to the average state in a well-defined manner.

The meaning of `sufficiently complete system policy' is now clear. The
covering of the policy domain must be as large as the domain of state
one wishes to cover, since it follows from the above definitions that
the association of policies to states is now one to one, after one
factors out the equivalences $\cal G$.  The uniqueness is secured by
making the configuration instruction itself a part of the
state. Without this, there would still be ambiguity, since there is no
guarantee that a measurement $\overline M(C_r,C_u)$ is
a unique function of its arguments. This completes the proof.

This sufficiency referred to above has the corollary that an
incomplete system policy $P_1$ cannot determine a unique state for the
whole system, only a part of it. An incomplete policy divides the
system into two or more parts, since the total policy is still in one
to one correspondence with the states.
\beq
P_1+P_2+\ldots \rightarrow S_{p1} + S_{p2}+ \ldots 
\eeq
By the virtue of the fundamental theorem, we have the important
conclusion that the necessary and sufficient condition for
implementation a policy $P(t)$  (i.e. the ability to map it onto a system
configuration over a period of time) is that the total average state
$\overline S(t) = \overline S_p(t)$.

Let us take a moment to understand the structure and meaning of
average ideal state.  It is tempting to think of the system as being
in an ideal state at some time $t_0$ and then deviating from it at
later times.  The precise state of the system at some reference time
might seem to characterize an ideal to our subjective judgement, but
the ideal state of configuration must change with time, since the
computer system is, by nature, influenced by users whose activities
are not completely secured by a policy. To freeze one's view of the
ideal in time, is to place unreasonable restrictions on the use of the
system (we shall see this later in examples connected to the use of
fixed disk quotas).  A specification of resource and user boundary
conditions is not the same as a specification of the ideal dynamical
behaviour of the system, if users are allowed to act on the resources.

Given that the policy and user configurations are stable over the
prescribed time-scales, one may take the average value (or
distribution of values) for each metric which characterizes the
response of system over shorter time-scales $\overline M(C_r,C_u)$ as
being representative of the state at time $t$. This summarizes the
effect of feedback of users on resources, or the statistical
interaction between the users and the system. Since we have prescribed
every bit string affecting the dynamics of the system at the outset as
policy, and we have measured the average result of those prescriptions
at time $t$, we have a complete description of the system in terms of
an implementable policy.
\beq
\overline S_p = P \oplus \overline M(P).
\eeq
Not surprisingly, this expression is directly analogous to linear
response theory in the physics of time-varying systems.  The policy
plays the role of a constraint of the motion, while the statistical metric
$\overline M$, has the role of the integrated response of the evolved
state at time $t$. The `equations of motion' which lead to the
evolution of a system also have an analogue here: they are the
operations carried out by the {\em system software} on the resources.

\begin{figure}[ht]
\psfig{file=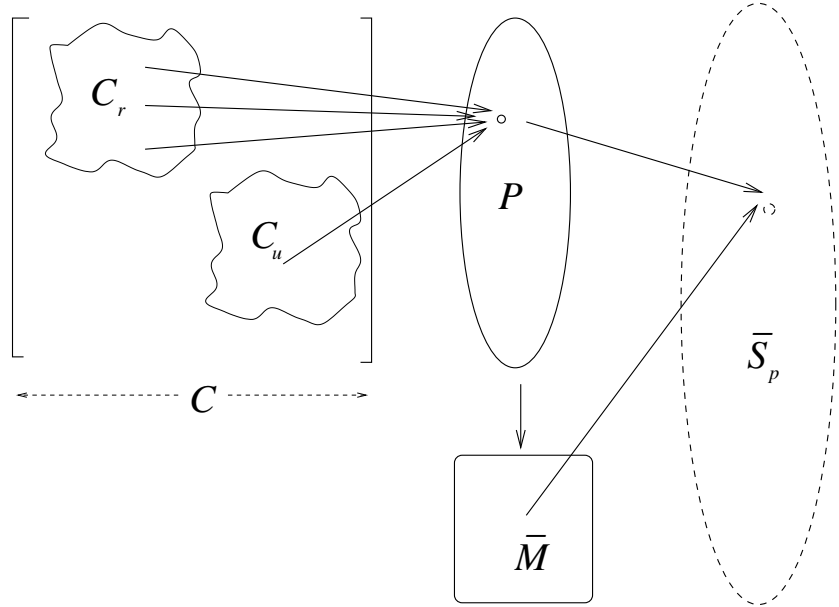,width=11cm}
\caption{The existence of an ideal state. This picture shows the mapping of
equivalent configurations of users $C_u$ plus hardware $C_r$ to a
factored set $C/{\cal G}$ which can be interpreted as the set of
implementable policies $P$. A unique configuration results in a
measurable effect on the system $\overline M$, i.e.
the feedback resulting from the policy $P$. The combination of the configuration
with its average effect on the system defines a unique state
$\overline S$.}
\end{figure}

The permutation or invariance group ${\cal G}$ is of no concern to
this paper except as a matter of principle for the most pedantic.  It
is a heuristic representation of all of the involved details which are
irrelevant to a theoretical formulation, but which occur in
practice\footnote{The fact that such details do indeed represent a
set, indeed a group of permutations and transformations is clear from
the empirical facts. The factoring of redundancy means picking only
one representative member of each configuration which gives the same
results, in the same manner that factor groups are formed in group
theory.  Indeed, it is a trivial technical point that the sets and
equivalence classes in a computer system may all be represented by
operations on a single binary string (a computing machine does
precisely this on a finite, possibly disjointed binary string). The
existence of transformations with closure is assured by extending a
binary string to encompass all possibiluties; the existence of an
inverse is trivial for permutations, as is the existence of a null
operation and associativity. That the formal factor group exists
follows from the existence of heuristic equivalences with respect to
system function.}.  Nevertheless, it has a theoretical implication: the ideal
state $\overline S$ has a number of equivalent representations,
e.g. those formed by permuting or swapping configuration details whose
ordering or equivalence is unimportant. This multiplicity could be a
benefit or a hazard to the task of implementation of policy. This remains to
be determined.

Having established the existence of an ideal state, the second
basic assumption is:
\begin{axiom}
{\it The long term aim of system administration is to optimize the
policy $P(t)$ for maximum productivity, insofar as this is allowed by
local constraints.  The short term aim is to keep the system as close
to the resulting ideal state $\overline S_p(t)$ as possible, i.e. to
minimize fluctuations $\delta S = \overline S - S$.}
\end{axiom}

One expects the average state $\overline S_p$ to exhibit {\em persistent}
behaviour however, i.e. be invariant for periods approaching the duration
of the system policy.
Note that, what we are essentially doing, by making the assertion of
an ideal state, is to separate slowly varying changes from quickly
varying changes.
\begin{equation}
S(t) = \overline S_p(t) + \delta S(t).
\end{equation}
Errors and misconfigurations (fluctuations $\delta S$) can accumulate
over short periods of time, shorter than the time scale over which the
average or ideal state changes. In terms of the relative rates of change:
\begin{equation}
\max\left|\frac{1}{\delta S}\frac{d }{dt} \delta S\right|\gg\max\left|
\frac{1}{\overline S_p}\frac{d}{dt} \overline S_p\right|.
\end{equation}
The business of system administration is therefore a problem in {\em regulation},
or in minimizing the effect of $\delta S$.

We arrive at the following: a theory of system administration would
attempt to answer the questions: 
\begin{quote}
{\em Is there an optimal strategy for
keeping the system as close as possible to its ideal state, and
maximize its productivity?}
\end{quote}
To answer these questions, we must understand more deeply the meaning of
the abstract formulation above. To begin, we backtrack and re-examine
the underpinning concepts.

\section{A generic computer system}

In order to elucidate the goals of computer configuration and maintenance, it will
be necessary to identify the main characteristics of computer systems at a
suitable level of abstraction. These include finding:
\begin{itemize}
\item Relevant variables,
\item Invariances,
\item Persistent structures,
\item Sources of information loss or entropy,
\end{itemize}
which affect the principal goals. Several studies of computer systems have
attempted to identify such
qualities\cite{huberman1,burgessLISA99,burgessPRL} and it is
hypothesized that a suitably abstracted description can be built on
the few simple principles identified by these authors.

The basic model of a computer is that of a dynamical community of
processes and resources, coupled to an external environment (an
external {\em source} or force). The source includes the stochastic
influences of all of the users of the system, and any other computer
systems which communicate with hosts within the perimeter of our own
system.  As pointed out in ref. \cite{NASAsteve}, the issue of
networking does not increase the complexity of the administration
problem, only its localization and perhaps its magnitude.  A set of
networked hosts, sending external messages, is simply a single virtual
host with internal inter-process communication.

The variables, important in characterizing the usage of a computer system,
are measures of average behaviour, such as rate of work, numbers of
processes, network connections and so forth\cite{burgessLISA99}. Other measures,
such as average service latencies, affect the system only at the level
of the network. Latencies are very complex phenomena and are unlikely to 
be predictable by any simple model.

Invariances refer to the independence of qualities and values to
changes. In the long run, there are no features of a computer system
which are fully constant, but for long periods of time, certain things
can be considered invariant.  For instance, the software tools one
uses to edit a file usually make no difference to the outcome, thus
the outcome of an editing operation may be considered invariant with
respect to differences in software used; the CPU efficiency of the
software used makes no difference to the result in most
cases. Invariance could also mean that a particular piece of software
never changes (is never upgraded), or that the content of a
configuration file is fixed with respect to other changes. In the
space of changes, such invariances may be considered to be {\em
ignorable coordinates}.

Persistent structures are, like invariances, values or qualities which
do not change over appreciable periods of time. This includes checksums
of important software, kernel profiles of software; it might also include
numbers of user accounts. Persistent structures are not expected to change.
Changes in these structures might be considered anomalous behaviour.

An important characteristic of computer systems is that they are
strongly coupled to human users' behaviour patterns. The majority of
human users follow strict daily and weekly work patterns and this is
reflected in many measurements of system resource behaviour. A
consequence of this is that measurements which are periodically
constrained are distributed according to a Planck spectrum. The Planck
spectrum can therefore be considered a general characteristic of computer
statistics in many cases. 

\section{The scope of a theory of system administration}

Even a limited theory of system administration should cover some
key aspects of the problem:
\begin{itemize}
\item Policy determination,
\item Strategic decisions about resource usage,
\item Productivity considerations (the economics of the system),
\item Empirical verification of strategies and policies,
\item Efficiency of policy and of policy implementation,
\item Efficiency of the system in doing its job.
\end{itemize}
More pragmatic details such as the need for software installation and
upgrade have to be tackled at an abstract level, in terms of productivity,
probability of failure, resource usage and so on. Software bugs can be
addressed in terms of productivity or security. Security, in turn can be
viewed as a contest for resources at the level of the system.

The benefits of automation versus human incursion are often discussed
in system administration, sometimes as a bone of contention. This is
one area that a theory of system administration can address objectively
and have a real prospect of answering once and for all. An aspect of this
will be discussed later as an example. 

\subsection{Measures and characters}

As an empirical science, system administration suffers from many
shortcomings. It has all of the problems associated with the social
sciences: statistical measures are seldom forthcoming, experimental
repeatability is a luxury, and sufficient repetition to obtain
statistically meaningful samples is a near impossibility. The
conditions under which measurements are made are constantly
changing. The situation is somewhat analogous to that of
non-equilibrium statistical mechanics in physics, but markedly less
controlled.

The characteristics which are of interest to us refer to the actions
and results which inter-weave in the dynamical behaviour of the system.
These include the quality of actions of the system administrator, in
relation to the prescribed policy, a typical characterization of the
environment which affects the system.  The measurements which are most
useful are those based on persistent variables, since these have a
stable value. Other fluctuating values can be treated stochastically
or averaged out into persistent values.

The following measures will be useful in formulating `pay-off' matrices
for administration models, as in the example to follow below.
The accuracy with which a policy is implemented by an agent of system
management (human or automatic system) can be gauged with the
following ratio:
\begin{eqnarray}
{\rm Accuracy} = \frac{\rm Number~ of~ policy ~actions}{\rm All~ actions performed}
\end{eqnarray}
i.e. the fraction of work which is within prescribed guidelines.
In algebraic terms:
\begin{eqnarray}
\alpha = \frac{N_p(t)}{N(t)} = \frac{\displaystyle \sum_{(a\subset P)} N_a}{\displaystyle \sum_{(\forall a)} N_a}
\end{eqnarray}
For humans $\alpha \leq 1$. For any bug-free automatic system, $\alpha = 1$.
Similarly, one may define the efficiency of a system by its use of resources (memory
and CPU share):
\begin{eqnarray}
{\rm Efficiency} = {\rm Accuracy} \times \left( 1 - \frac{\rm Resources~used}{\rm Resources~available}\right)
\end{eqnarray}
In algebraic terms:
\begin{eqnarray}
\varepsilon = \alpha\left(1 - \frac{\displaystyle \sum_{(a\subset P)} r_a}{\displaystyle \sum_{(\forall a)} r_a}\right)
\end{eqnarray}
i.e. the more resources which are consumed in implementing a policy, the less
efficient it can be considered to be.

Other measures are more useful for describing the relationship of a
computer system to its environment, or the influential forces which
steer its dynamical evolution.  The response of a computer system to
its users is characterized by averages which fluctuate in time. Human
society's diurnal work pattern imposes a twenty four hour periodic
character on these measurements\cite{huberman1,burgessLISA99} and a
also a weekly work pattern, which is dominant during weekdays and
slight at weekends (at least in the Western world). The periodic
topology implies that the distribution of resource usage takes
on the special form of a Planck distribution with a Gaussian component,
by analogy with statistical physics at temperature $T$:
\beq
D(\lambda) = A\;e^{-\left(\frac{(\lambda-\overline\lambda)^2}{2\sigma^2}\right)} + 
\frac{B}{(\lambda-\lambda_0)^3(e^{1/(\lambda-\lambda_0) T}-1)}.
\eeq 
$\lambda$ is the deviation of a measurement from its
average value over a period.  The values of the constants $A$, $B$, $\lambda_0$
and $T$ may be chosen to fit the behaviour of any variable which is strongly
coupled to periodic usage. Their absolute values have no significance,
since there is no `standard candle' computer system to compare to, but
changes relative to the local norm could be interpreted as anomalies.
Non-zero $A$ allows for the presence of additional Gaussian noise in
some measurements.

\subsection{Interactions of time scales}

The identification of suitable time-scales is of crucial importance
to any dynamical problem. Time-scales control rates of competition which
lead to balance, and also rates of change.

It is easy to show that human administrators only compete with
automatic systems in speed and efficiency at times of the day when
they have nothing pressing to do. Indeed, it is always possible to
arrange for an automatic system to beat a human, provided it can run
in overlapping instantiations.  A straightforward comparison of the
time-scales involved in automated maintenance, to those of manual human
maintenance can be made for any operation which is programmable in an
automatic system with available technology.

Alarm systems which merely notify humans of errors and then rely on a
human response are intrinsically slower than automatic systems which
repair errors, provided the alarms represent errors which can be
corrected with current automation.

The response time $t_{\rm auto}$ of a automatic machine system $M$,
falls between two bounds (see figure 1)
\begin{eqnarray}
n\, T_p +T_e(A) ~\geq ~ t_{\rm auto}~ \geq ~ T_e(A)
\end{eqnarray}
where $T_p$ is the scheduling period for regular execution of the
system (e.g. the cron interval, typically half-hour to an hour),
$T_e(A)$ is the execution time of the automatic system (typically
seconds). The integer $n \ge 0$ since the number of iterations of the
automatic system required to fix a problem might be greater than
one. The time required to make a decision about the correct course of
action $T_d(A)$ is negligible for the automatic system. 

\begin{figure}[ht]
\psfig{file=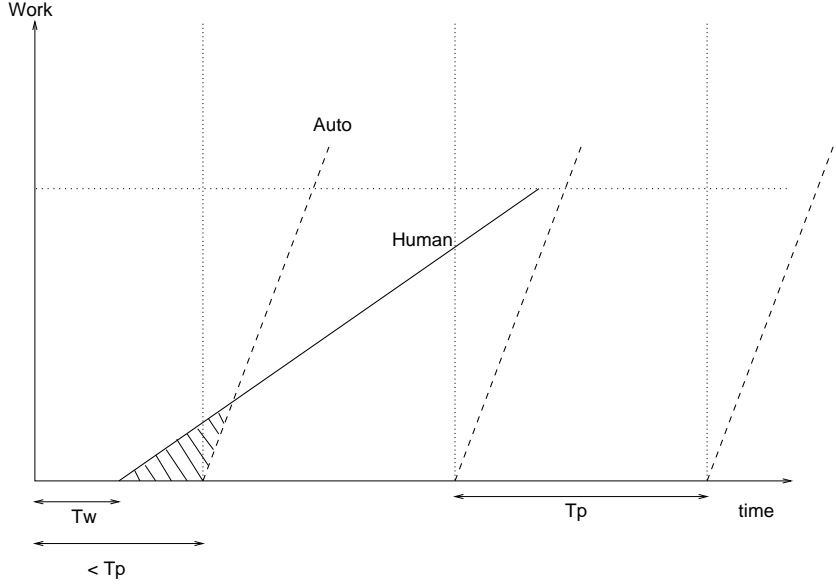,width=11cm}
\caption{Overlapping work rates of human and automatic systems.}
\end{figure}

For a human being, making a decision based on a predecided policy, the
response time $ t_{\rm human}$ falls between the limits:
\begin{eqnarray}
\infty \geq t_{\rm human} \geq T_w(H) + T_d(H) + T_e(H). 
\end{eqnarray}
$T_d(H)$ is again the decision time, or time required to determine the
correct policy response (typically seconds to minutes).  $T_e(H)$ is
the time required for a human to execute the required remedy
(typically seconds to minutes).  $T_w(H)$ is the time for which the
human is busy or unavailable to respond to the request, i.e. the
wait-time. The availability of human beings is limited by social and
physiological considerations. In a simple way, one can expect this to
follow a pattern in which the response time is greatest during
the night; simplistically, if one assumes that humans sleep 8 hours,
\begin{equation}
T_w(H) > 4(1 +\sin(t/24)),
\end{equation}
where time is measured in hours, whereas
\beq
T_w(A) \simeq 0.
\eeq
We can note that human response times are usually much longer than the
corresponding machine response times,
\begin{eqnarray}
T_d(A) \ll T_d(H)\nonumber\\
T_e(A) \ll T_e(H)
\end{eqnarray}
and that the periodic interval of execution of the automatic system is
generally taken to be greater than the execution time of the
automatic system
\begin{eqnarray}
T_p \geq T_e,
\end{eqnarray}
thus avoiding overlapping executions (though this is not necessarily a
problem, see the discussion of adaptive locks\cite{burgess3}) It is
always possible to choose the scheduling interval to be arbitrarily
close to $T_e(A)$ (i.e. as short as one likes). Then provided,
\beq
T_w(H) > T_e(A)
\eeq
the automatic system can always win over a human.  This last
inequality requires qualification however, since very long jobs (such
as backups or file tree parses) increase exponentially in time with
the size of the file tree concerned. This makes a prediction: it tells
us that one should always arrange to allow such long jobs to be run
last in a sequence of maintenance tasks, and also in overlappable
threads. This means that long jobs will not hinder the rapid execution
of a maintenance program.

Cfengine\cite{burgess1} allows overlapping runs using its scheme of
adaptive locks\cite{burgess3}.  Thus, by scheduling long jobs last in
a cfengine program, it is {\em virtually} always possible for cfengine to beat a
human, unless it is prevented from running, or the human is given the
chance to respond with a head-start; this seldom happens by chance.

\section{Primitive moves}

Having identified the principle aims and methods of system
administration, one is free to represent a model for these in any convenient
calculational scheme. Almost immediately, one is confounded by the
multiplicity, or non-uniqueness of the mapping between problem and
solution: It is common-lore amongst system administrators, and it is to
be expected logically in any causal web, that
\begin{itemize}
\item One problem can have several solutions.
\item Several problems can be solved with a single solution.
\end{itemize}
How should one classify such mappings? By coarse-graining? Some degree
of coarse classification is inevitable to make the analysis tractable,
but it needs to be performed in a well-defined way. To some extent, we
have already dealt with this problem in the factoring out of redundant
expression in section 2. However the same problem returns in
specifying the actions required to maintain the ideal state.

In order to unravel this situation as far as possible, it is
reasonable to try to express problems and solutions in terms of linear
combinations of primitive actions. The analysis of primitive operations has
already been considered by Burgess in ref. \cite{burgess1} in
developing automated approach to system administration. There is little
to add here, except to say that it is required that every implementable policy be
decomposable as a combination of these primitives. 

The available channels for action, i.e. the possible moves which a
`player of the computer system game' (user or administrator) can
choose from, form a huge set if one views them at the level of the
user. Formulating generic activity would be an intractable problem if
one chose to consider every nuance of the system, viewed from a user
perspective. Fortunately it is possible to break down the variety of
activities available to users into a number of primitive actions. Any
task can be considered as some linear combination of these few basic
actions. The actions are:

\begin{center}
\begin{tabular}{|l|l|}
\hline
Primitive type $T^i$ & Comments/Examples\\
\hline
Create file & \\
Delete file & Tidy garbage\\
Rename file & Disable\\
Edit file & Used in configuration\\
Access control & Permissions\\
Request resource & Read/Mount\\
Copy file & Read/write\\
Process control & Start/stop\\
Process priority & Nice\\
Configure device & \\
\hline
\end{tabular}
\end{center}

We should be careful to distinguish between how functions are implemented
and how they can be decomposed. The method of implementation is not
necessarily relevant to the analysis. What is important is that there exists
a finite number of primitive actions which can be used to express all others
in combinations.

Are these primitives sufficient in themselves? Could we implement the
following policy, for instance: downloading of pornographic material
between the hours of 9:00 and 17:00 is forbidden? If such a policy is
implementable by an automatic system, it must be possible to filter
content-specific data.  Such a filter would need a configuration file
which would need to be edited.  The time-dependent behaviour could be
handled by a scheduler, also configured by a text file. These
configuration details are all implementable with file editing and
process control. The ability of software to perform the task has to be
assumed. This has nothing to do with management of the system.
 If the same job is to be carried out by a human, then the
model of the management system must be extended to include humans, in
which case job control and job definition require the analogous
concepts to file editing and process control, for human brains. In
other words, when humans are involved in a theory of manual work, they must be
considered a part of the computer system.

\section{The ideal average state}

In order to have a chance of repairing damage, or maintaining a
detailed balance of resources, we need to be able to trace the
development or history of the system, from an ideal average state at
an initial time, to a less than ideal state at a later time.  In
accordance with the axioms lain out at the beginning of the paper, it
is assumed that the ideal state is determined as a matter of policy,
by local considerations.

Many minor changes take place all the time in a computer system; these
are healthy. Programs are started and stopped, files are created and
destroyed: this is part of the work done by the system. However,
certain features of the system should not change greatly (they should be
persistent, at least on average). For instance, resources like disks
and network services should be available to users at all times.
If a crucial service falls out, then it affects other changes in the
system.

Some changes are important to operation of the system, others are
unimportant.  For instance, it would be unimportant if one swapped the
process ID's of two programs. The process ID is just a label which has
no bearing on the performance of the system or the productivity of
users. However, if one process stopped running prematurely, this would
be a change of state.

If we know what changes have taken place to move the system away from
the ideal state, it should be possible to undo them, provided these do
not involve the destruction of useful work. To accomplish this
tracking of changes, in formal terms, we need to quantify the state of
the system with respect to specific changes\footnote{Note that, while
one is interested in tracking changes in principle, in order to
formulate the theory of system changes, this does not imply memorizing
changes in a system is a desirable thing to do. Some system
administration tools attempt to do this, often unsuccessfully, but as
a counter-example one has cfengine which simply acts as a generic
counter-force, pushing the system towards the ideal state, regardless
of what specific historical chain the system follow.}.  Suppose one
considers system administration as a game, framed on a lattice of
$n$-dimensions, and suppose that the system has an ideal state located
at the origin of this lattice, based on a policy and described in
terms of primitive system variables. Each node of the lattice is a new
state of the system.  Let us suppose that the aim of the game is to
remain as close as possible to the ideal state, i.e. the origin of
this discrete space.  How can one formulate such a game? How many
dimensions does the lattice extend into, and what do they represent?
These questions are central to formulating an analysis.

In a general sense, a computer system is a dynamical system like any other,
and it must follow the same basic principles as any set of variables which
changes in time. 
Let $\phi_i(t)$, where $i=1,2,...N$ be the set of measurable variables
which can be associated with a computer system. A canonically complete
dynamical system can be associated with the set of phase-space variables,
$$
q_i(t)~,~\dot q_i(t),
$$
i.e. the variables and their time derivatives.
Not all variables can be considered differentiable functions of time, but
it will be possible to give the derivative a meaning even for discrete
variables, so this may be regarded symbolically for the present.
Given that the values of these variables can change statistically with time
(the nature of this variation will be qualified later),
at any time $t$, we can decompose the value of $q(t)$ into a local
average and a fluctuating piece.
\beq
q(t) = \overline q(t) + \delta q(t).
\eeq
This means essentially decomposing $q(t)$ into fast and slowly
changing variables.  The average value $\overline q(t)$ varies only
slowly with time, but many rapid changes $\delta q(t)$ fluctuate about
the average value. The average may be defined by
\beq
\overline q_i(t) = \frac{1}{t-t_i} \int_{t_i}^{t}\;q_i(t')\,dt',
\eeq
where $t-t_i$ is the interval over which the average is taken,
and it is assumed that
\beq
t - t_i < t - t_0,
\eeq
where $t_0$ is the `zeroth' time at which the system was in the ideal
state. The rate at which variables are changing $\dot q(t)$ can also
be measured.  A similar procedure can be implemented for the $N$
derivatives and their local average values.

For the sake of characterizing the state of the system, one is
interested in change in the average values since some ideal zeroth
time $t_0$:
\beq
d_i \equiv \left\{ \overline q_i(t) - \overline q_i(t_0), \overline{\dot{q_i}}(t) - \overline{\dot{q_i}}(t_0) \right\}.
\eeq
In terms of the deviations $d_i$ in key system variables, one may
postulate a $2N$-dimensional lattice whose independent, orthogonal
axes are the $n=2N$ variables of the phase space $d_i$, for $i=1...n$.
Positions on this lattice are denoted by the vector of these component
deviations. It is collectively denoted $\vec d$.

Suppose now that the system has deviated from the ideal state at $\vec
0$ and has reached a point $\vec d$ on the lattice (see figure 2).  The number of
equivalent paths $H(\vec d)$ back to the ideal state, is
\begin{eqnarray}
H(\vec d) ~~=~~ 
\frac{{\displaystyle (\sum_{j=1}^n d_j)!}}{{\displaystyle \prod_{k=1}^n (d_k!)}}
\end{eqnarray}
This grows rapidly with the Euclidean distance $|\vec d|$ 
\beq
|\vec d| \equiv d = \sqrt{\displaystyle{\sum_{i=1}^n (d_i)^2}}.
\eeq
$H(\vec d)$ may be considered as a measure of the entropy,
or disorder in the system.  The entropy may be thought of as measuring
the `hopelessness' of finding the original route which led to the
deviation. If all the paths are equivalent, i.e. the particular route
by which the current state was achieved was not important, then it
measures the number of equivalent ways in which the deviation can be
fixed.
\begin{figure}
\psfig{file=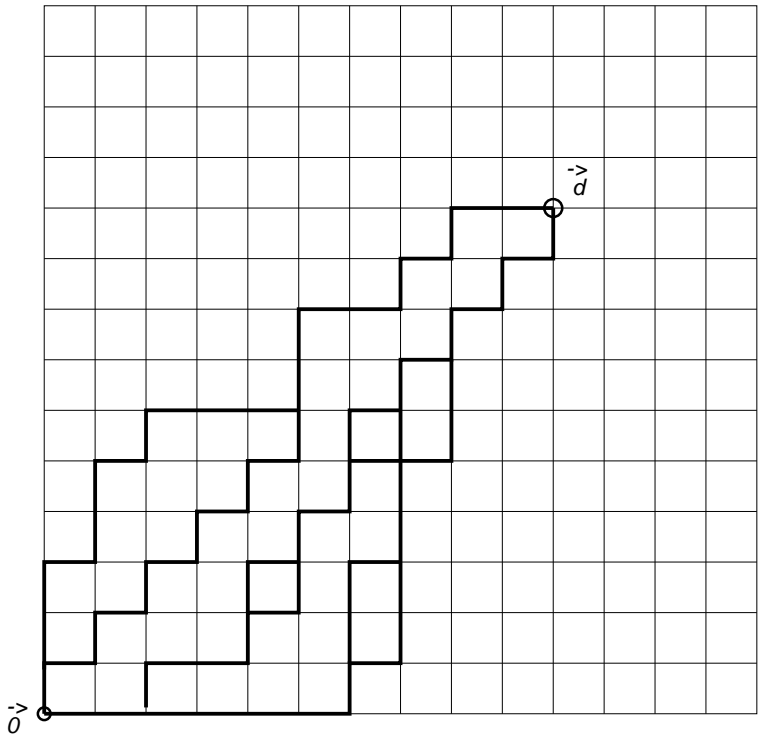,width=10cm}
\caption{Deviations from the ideal state may be visualized as a random
walk through a lattice of $n$-dimensions (here only two). The number
of paths of equal length by which one can return to the origin increases
rapidly with the distance.}
\end{figure}

If the path is important then a different interpretation is more
appropriate.  In common with its analogue from physics, $H$ may be
thought of as a measure of the amount of potential work has been lost
to the system as a result of its deviation from the ideal state. Or
conversely, here it may be considered a measure of the amount of work
which would have to be expended in order to return the system to its
ideal state. To gauge how quickly this grows with distance, one may compute
the rate of increase in numbers of paths as $\vec d$ increases.
Define
\begin{eqnarray}
\stackrel{d_i}{\nabla} H(\vec d) &=& \frac{H(\vec d + \vec {\Delta d})-H(\vec d)}{|\vec{\Delta d}|}\nonumber\\
&=& H(d_1,..,d_i+1,..d_n) - H(d_1,..,d_i,..d_n)
\end{eqnarray}
Thus we define the rate of increase on the discrete lattice by,
\begin{eqnarray}
\frac{\stackrel{d_i}{\nabla} H(\vec d) }{ H(\vec d)} = \frac{1}{(d_i + 1)}\;\sum_{j\not=i}^n d_j \gg 1.
\end{eqnarray}
This shows that the increase is in fact approximately
proportional to the distance. In other words, the rate of increase is
approximately exponential. Clearly, this simple quantification of cumulative
system error indicates that deviations from an ideal state should be dealt
with as quickly as possible, since it becomes increasingly difficult to
make corrections as the errors are compounded.

The ideal state itself needs to be characterized in terms of
reasonable tolerances in system variables. The important variables
include the availability of resources (ability to create new files and
processes) as well as the level of activity. In dynamical terms one
considers a set of variables and their rates of change:
\beq
q(t)~,~ \frac{dq(t)}{dt}.
\eeq
If the system is a complete characterization of every possible influence
and change in the system, then these form a simplectic algebra and the behaviour
of the system is, at least in principle, completely deterministic, if not exactly
predictable. In most cases there are influences which are not completely known,
or may be regarded as random. In that case, one moves from simple mechanical
systems into to realm of statistical mechanics and non-equilibrium studies.

These underlying variables are only indirectly linked to the ideal state,
through averaging.

The above view is quite simplistic. In reality there might not be only
one ideal state, but a set of equivalent ideal states. These can all
be formulated as direct sums or quotients of a simply-connected state
space however, so these need not be of concern to the principle of the
argument. Having identified an ideal state as a point in a vector space, or
lattice, one is now free to discuss how changes in the forces or influences
on the system lead to movements through the lattice.

\section{Game theory and the contest for the ideal state}

There are two separable issues in the ideal-state view of system
administration. The distinction concerns the perceived
intelligence behind the changes which lead to a degradation of the
ideal state. We may classify changes as either random (stochastic)
or as intentional (strategic) depending on the nature of the adversary.

This distinction is partly artificial: all changes can be traced back
to the actions of humans at some level, but it is not always pertinent
to do so.  Not all users act in response to a specific provocation, or
with a specific aim in mind. It just happens that their actions lead
to a general degradation of the ideal state, no malice intended.  This
strikes back to the fundamental principle of detail, namely that high
level effects wash out the specifics of low-level origins. Thus there
is a part of the spectrum of changes which averages out to a kind of
faceless background noise. The details of who did what are of no
concern. Random influences have been analyzed in
ref. \cite{burgessLISA99} and are found to follow a number of
well-known statistical distributions. Their study is part of the problem
to be solved, but not all of it.

The other part of the problem is the case of actions which may be
regarded as being more carefully calculated, or following a
systematic behavioural pattern. These are caused by conflicts of
interest between system policy and user wishes.  A suitable framework
for analyzing conflicts of interest, in a closed system, is the theory
of games\cite{morgenstern,dresher}.  
Game theory is about introducing players, with goals and
aims, into a scheme of rules and then analyzing how much a player can
win, according to those restrictions. Each move in a game affords the
player a characteristic value, often referred to as the `payoff'.
Game theory has been applied to warfare, to economics (commercial
warfare) and many other situations. In this case, the game takes place
on the $n$-dimensional board, spanned by the $\vec d$ vectors.

There are many types or classifications of game. Some games are
trivial: one-person games of chance, for example, are not analyzable
in terms of strategies, since the actions of the player are irrelevant
to the outcome. In a sense, these are related to the first kind of
deviation referred to above. Some situations in system administration
fit this scenario.  More interesting, is the case in which the outcome
of the game can be determined by a specific choice of strategy on the
part of the players.  The most basic model for such a game is that of
a two-person zero-sum game, or a game in which there are two players,
and where the losses of one player are the gains of the other. `Zero
sum' is the law of conservation of currency (current).

Many games can be stated in terms of this basic model, although this
is often a simplification of reality. Games in complex systems are
rarely true zero-sum games: energy leaks out, money gets burned or
printed and thus there is no exact zero-sum conservation.

\subsection{Models}

The basic valuables of system administration are the system resources:
file space, CPU share, memory share and network share. The theory of
system administration can be viewed as a competition for these
resources and for user privileges. The central obstacle in formulating
a scenario in terms of game theory is the classification of strategies
and their evaluation in terms of a characteristic (payoff) matrix.

\begin{itemize}
\item As a {\em zero sum, two person game} system administration is a game between
the collective users and the system administrator. The aim of the
users is to consume all of the system resources, while the aim of the
administrator is to keep the system as close as possible to its ideal
state. Ideally, the system administrators strategies should always
bring the system closer to the ideal state. This is the property of
{\em convergence} referred to in ref. \cite{burgess1,burgessLISA98}.
The ideal outcome of this game is a stalemate, or equilibrium
somewhere close to the ideal state. 

This game is often one with {\em perfect information} since all the
important moves are visible to both players, however both sides can
engage in bluffing. Clearly the administrator can win, either by
limiting or reducing the consumption of resources and by extending the
resources of the system. A user can `win' in a certain pessimistic
sense by moving the ideal state so far from the ideal that the system
crashes and thus the game ends.

\item A more optimistic variant of the above,
is to view the aim of users as being to produce as much useful work as
possible. This is a more complicated aim, since users can now impede
their own progress by consuming too many resources, thus impairing the
system as a whole and preventing themselves from being able to work
(users need to be environmentally friendly). Experience from reality
shows that most users do not concern themselves with this aspect
however; they see it as the system administrators job to deal with
such problems when they arise.

\item As a {zero sum, $N$-person game} one can make a more detailed model,
in which users compete against one another in addition to the system
administrator. The system administrator's task then becomes to act as
a kind of Robin Hood character, preventing any one user's consumption
of all resources, trying to distribute resources fairly. Again, the
aim of the administrator is to maximize the duration of the game by
keeping the system as close to the ideal state as possible.

\end{itemize}

\subsection{Payoffs and work}

The next obstacle concerns the level at which we decide to address
the behaviour of the system. Appropriate measures can be defined 
at various levels.

In order to formulate the characteristic matrix (often called the
pay-off matrix) we must identify the book-keeping parameters and aims
by which one hopes to win the game. What is the currency of this
system? In social systems one has money as the book-keeping parameter
for transactions. In physical systems, one has energy as the
book-keeping parameter. These quantities count resources, in some
well-defined sense. An analogous quantity is needed in system
administration.

\begin{itemize}
\item The aim of the system administrator is to keep the system
alive and running so that users can perform useful work.

\item The aim of benign users is to produce useful work using the
system. The aim of malicious users is often to maximize their
control over system resources.
\end{itemize}
In a community, games are not necessarily cut and dried zero-sum
engagements. We are faced with a Nash problem, or prisoner's dilemma,
which often ends in a Nash equilibrium\cite{nash1}.
\begin{quote}
A user of the system who pursues solely private interests,
does not necessarily promote the best interest of the
community as a whole.
\end{quote}
In other words, users can shoot themselves in the proverbial foot by using up all the
available resources on a finite system. 
This affects them as much as anyone else. The empirical evidence suggests that, on
average, users consume resources at a rate which is periodic and
polynomial in time\cite{huberman1,burgessLISA99}.
\begin{eqnarray}
W(t) \propto \sin(\Omega t)\,\sum_n c_n t^n.
\end{eqnarray}
A definition of work is required in order to quantify the production
of useful work in a non-prejudicial manner. Clearly the term `useful
work' spans a wide variety of activities. Clearly work can increase
and decrease (work can be lost through accidents), but this is not
really germane to the problem at hand. The work generated by a user
(physical and mental work and then computationally assisted results)
is a function of the information input into the system by the user.
Since the amount of computation resulting from a single input might
be infinite, in practice, the function is an unknown.

In general, the pay-off in not just a scalar value, but a vector. This
indicates that a game might actually be decomposable into a number of
parallel but interacting games.

What is the value of a game? How much can a user or an attacker hope
to win? The system administrator, or embodiment of system policy, is
not interested in winning the game, but rather in confounding the game
for users who gain too much control. The system administrator plays a
similar role to that of a police force. In some vague sense, the
administrator's jobs is to make sure that resources are distributed
fairly, according to the policies laid down for the computer society.

\subsection{Strategy expression}

In a realistic situation one expects both parties in the two-person game
to use mixed strategies. The formulation of the game theoretical pay-off matrix
requires one to consider the strategies which the players can adopt. Again, the
number of possible strategies is huge and the scope for strategic contrivance is
almost infinite. In order to limit the formulation of the problem, it is necessary
to break down strategies into linear combinations of primitives again.
What is a strategy? 
\begin{itemize}
\item A set of operations
\item A schedule of operations
\item Rules for counter-moves
\end{itemize}
In addition to simple strategies, there can be meta-strategies, or long-term goals.
For instance, a nominal community strategy might be to:
\begin{itemize}
\item Maximize productivity or generation of work.
\item Gain the largest feasible share of resources.
\end{itemize}
An attack strategy might be to
\begin{itemize}
\item Consume as many resources as possible.
\item Destroy key resources.
\end{itemize}
Other strategies for attaining intermediate goals might include covert
strategies such as {\em bluffing} (falsely naming files).  Defensive
strategies might involve taking out an attacker, counter attacking, or
evasion (concealment), exploitation, trickery, antagonization,
incessant complaint (spam), revenge etc. Security and privilege,
levels of access, integrity and trust must be woven into algebraic
measures for the pay-off.  A means of expressing these devices must be
formulated within a language which can be understood by system
administrators, but which is primitive enough to enable the problem to
be analyzed in an unambiguous fashion.

\subsection{Stable and dominant strategies}

It has been argued here, and in earlier
papers\cite{burgessLISA98,burgessLISA99}, that computer systems can be
viewed as fluctuating around statistically stable configurations, for
the most part. This assumes that both users and system administration
mechanisms are in approximate balance. Game theory is suited to
finding equilibria, or stable superiorities in a set of strategies.
Let us consider how game theory can be used to frame system behaviour
as a contest for control of the system's resources.

The simplest case of a two-person, zero-sum game is chosen.

We are interested in determining whether any optimal strategies can be
adopted by the system (and its administrator) in order to maintain
control of the system, i.e. in order to prevent users from winning
control of the system. This situation is analogous to the analysis of
dominant evolutionary strategies, considered by
Hamilton and Maynard-Smith\cite{maynardsmith}. These
so-called Evolutionary Stable Strategies are the winning
strategies favoured by natural selection mechanisms in the animal or
plant kingdom. In our case, we are simply interested in strategies
which are clear winners over all other strategies. If we consider the
characteristic matrix, or pay-off matrix, as a function of strategies
for attack and defense $\pi(\sigma_a,\sigma_d)$, then one may
characterize a dominant attack-strategy $\sigma_a^*$ by the criterion:
\beq
\pi(\sigma_a^*,\sigma_d) &>& \pi(\sigma_a,\sigma_d)
\eeq
i.e. $\sigma_a^*$ must be a better move than any other strategy
against and arbitrary counter-move $\sigma_d$.
If this is the case, then there exists at least one pure strategy which
is optimal for the attacker. Similarly, an optimal defensive strategy
$\sigma_d^*$ is characterized by:
\beq
\pi(\sigma_a,\sigma_d^*) &>& \pi(\sigma_a,\sigma_d)
\eeq
A more general situation is that one can find a winning mixture of
strategies $\Sigma$ (a linear combination of pure strategies)
\beq
\Sigma = \frac{1}{N}\sum_i^N c_i\,\sigma_i.
\eeq
Then if the dominant mixture of strategies $\Sigma_a^*$ satisfies,
\beq
\pi(\Sigma_a^*,\Sigma_d) &>& \pi(\Sigma_a,\Sigma_d)
\eeq
then the attacker must win, but if some optimal mixture of strategies
$\Sigma_d^*$ satisfies,
\beq
\pi(\Sigma_a,\Sigma_d^*) &>& \pi(\Sigma_a,\Sigma_d),
\eeq
then the defender must prevail. It is this final solution which one hopes to
find in order to secure a stable computer environment.

To illustrate this idea, consider an example of some importance, namely the
issue of garbage collection. The need for forced garbage collection
has been argued on several occasions\cite{lisa8941,burgess1,burgess3}, but the value of
this strategy to system rule has not been analyzed previously.

The first issue is to determine the currency of this game. What
payment will be transferred from one player to the other in play?
Here, there are three relevant measurements to take into account: (i)
the amount of resources consumed by the attacker (or freed by the defender),
and  sociological rewards: (ii) `goodwill' or (iii) `privilege' which are
conferred as a result of sticking to the policy rules. These latter
rewards can most easily be combined into an effective variable
`satisfaction'. Then the player who can't get no satisfaction is 
the poorer one. A satisfaction measure is needed in order to balance
the situation in which the system administrator prevents users from using
any resources at all. This is clearly not a defensible use of the system,
thus the system defenses should be penalized for restricting users
too much.
The characteristic matrix now has two contributions,
\beq
\pi = \pi_r({\rm resources}) + \pi_s({\rm satisfaction}).
\eeq
It is convenient to define
\beq
\pi_r \equiv \pi({\rm resources}) = \2\,\left(\frac{\rm Resources~won}
{\rm Total~resources}\right).
\eeq
Satisfaction $\pi_s$ is assigned arbitrarily from values from plus to minus one half,
such that,
\beq
-\2 \le \pi_r \le +\2\nonumber\\
-\2 \le \pi_s \le +\2\nonumber\\
-1  \le \pi \le +1.
\eeq
The pay-off is related to the movements made through the lattice $\vec d$.
The different strategies can now be regarded as duels, or games of timing.

\begin{center}
\begin{tabular}{|c|c|c|c|c|}
\hline
Users/System & Ask to tidy & Tidy by date & Tidy above & Quotas\\
             &             &              & Threshold  &       \\
\hline
Tidy when asked & $\pi(1,1)$ & $\pi(1,2)$ & $\pi(1,3)$ & $\pi(1,4)$\\
\hline
Never tidy & $\pi(2,1)$ & $\pi(2,2)$ & $\pi(2,3)$ & $\pi(2,4)$\\
\hline
Conceal files & $\pi(3,1)$ & $\pi(3,2)$ & $\pi(3,3)$ & $\pi(3,4)$\\
\hline
Change timestamps & $\pi(4,1)$ & $\pi(4,2)$ & $\pi(4,3)$ & $\pi(4,4)$\\
\hline
\end{tabular}
\end{center}
The elements of the characteristic matrix must now be modelled by suitable
algebraic or constant terms. The rate at which users produce
files may be written
\beq
r_u = \frac{n_br_b + n_gr_g}{n_b+n_g},
\eeq
where $r_b$ is the rate for bad users and $r_g$ is the rate for good
users.  The total number of users $n_u = n_b+n_g$. From the authors
experience, the ratio $n_b/n_g$ is about one percent. The rate can be
expressed as a scaled number between zero and one, for convenience, so
that $r_b = 1-r_g$.

The payoff in terms of the consumption of resources by users, to the users
themselves, is then
\beq
\pi_u = \2 \int_0^T dt\;\frac{r_u\,(\sin (2\pi t/24)+1)}{R_{\rm tot}},
\eeq
where the factor of 24 is the human daily rhythm, measured in hours,
and $R_{\rm tot}$ is the total amount of resources to be consumed.
Note that, by considering only good user or bad users, one has a
corresponding expression for $\pi_g$ and $\pi_b$, with $r_u$ replaced
by $r_g$ or $r_b$ respectively.  An automatic garbage collection
system results in a negative pay-off to users, i.e. a pay-off to the
system administrator. This may be written
\beq
\pi_a = -\2 \int_0^T dt\;\frac{r_a\,(\sin (2 \pi t/T_p)+1)}{R_{\rm tot}},
\eeq
where $T_p$ is the period of execution for the automatic system, considered
earlier. This is typically hourly or more often, so the frequency of the
automatic cycle is some twenty times greater than that of the human cycle.
The rate of resource-freeing $r_a$ is also greater than $r_u$, since file
deletion takes little time compared to file creation, and also an automated
system will be faster than a human. The quota payoff yields a fixed allocation
of resources, which are assumed to be distributed equally amongst users
and thus each quota slice assumed to be unavailable to other users. The
users are nonchalant, so $\pi_s=0$ here, but the quota yields
\beq
\pi_q = +\2\left( \frac{1}{n_b+n_g}\right).
\eeq
The matrix elements are expressed in terms
of these.

\begin{itemize}
\item[$\pi(1,1)$:] Here $\pi_s=-\2$ since the system administrator is maximally
satisfied by the users' behaviour. $\pi_r$ is the rate of file creation by
good users $\pi_g$, i.e. only legal files are produced. Comparing the strategies, it
is clear that $\pi(1,1)=\pi(1,2)=\pi(1,3)$.

\item[$\pi(1,4)$:] Here $\pi_s=0$ since the users are
dissatisfied by the quotas, but the system administrator must be
penalized for restricting the functionality of the system. With fixed
quotas, users cannot generate large temporary files.
$\pi_q$ is the fixed quota payoff, a fair slice of the resources.
Clearly $\pi(4,1)=\pi(4,2)=\pi(4,3)=\pi(4,4)$. This tells us that
quotas put a straight-jacket on the system. The game has a fixed
value if this strategy is adopted by system administrators. However,
it does not mean that this is the best strategy, according to the
rules of the game, since the system  administrator loses points for
restrictive practices. This is yet to be determined.

\item[$\pi(2,1)$:] Here $\pi_s=\2$ since the system administrator is maximally
dissatisfied with users' refusal to tidy their files. The pay-off for
users is also maximal in taking control of resources, since the system administrator
does nothing to prevent this, thus $\pi_r=\pi_u$. Examining the strategies,
one find that $\pi(2,1)=\pi(3,1)=\pi(3,2)=\pi(3,3)=\pi(4,1)=\pi(4,2)$.

\item[$\pi(2,2)$:] Here $\pi_s=\2$ since the system administrator is maximally
dissatisfied with users' refusal to tidy their files. The pay-off for
users is now mitigated by the action of the automatic system which works
in competition, thus $\pi_r=\pi_u-\pi_a$. The automatic system is invalidated
by user bluffing (file concealment).

\item[$\pi(2,3)$:] Here $\pi_s=\2$ since the system administrator is maximally
dissatisfied with users' refusal to tidy their files. The pay-off for
users is mitigated by the automatic system, but this does not activate until
some threshold time is reached, i.e. until $t>t_0$. Since changing the
date cannot conceal files from the automatic system, when they are
tidied above threshold, we have $\pi(2,3)=\pi(4,3)$.
\end{itemize}
\begin{figure}
\psfig{file=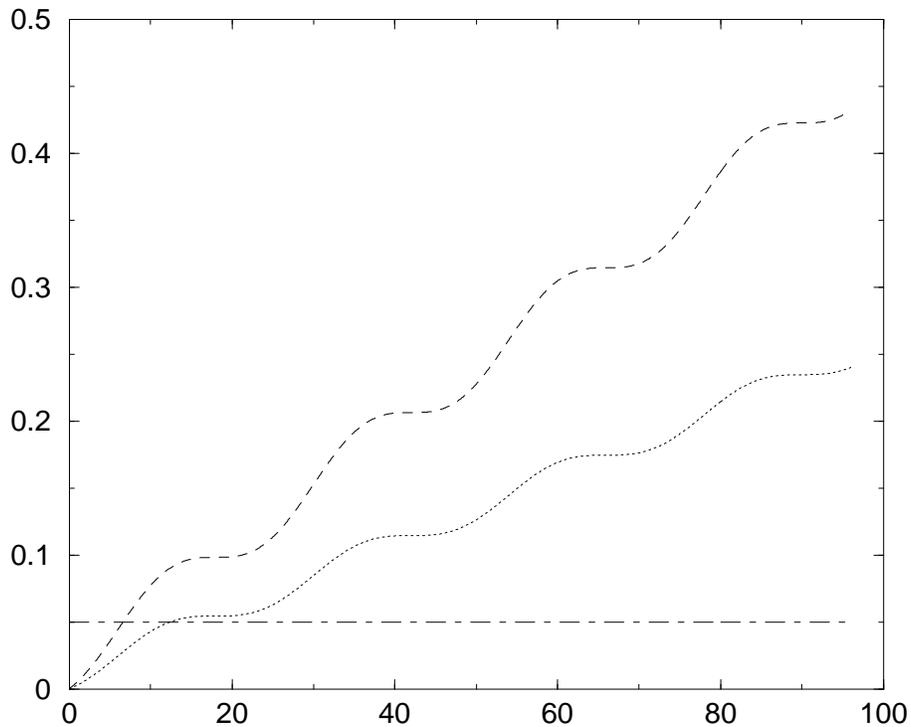,width=12cm}
\caption{The absolute values of pay-off contributions as a function of time (in hours),
For daily tidying $T_p=24$. User numbers are set in the ratio $(n_g,n_b)=(99,1)$,
based on rough ratios from the author's College environment, i.e. one percent
of users are considered mischievous. The filling rates are in the same ratio:
$r_b/R_{\rm tot}=0.99, r_g/R_{\rm tot}=0.01, r_a/R_{\rm tot}=0.1$. The flat
dot-slashed line is $|\pi_q|$, the quota pay-off. The lower wavy line is the
cumulative pay-off resulting from good users, while the upper line represents
the pay-off from bad users. The upper line doubles as the magnitude of the pay-off
$|\pi_a|\ge|\pi_u|$, if we apply the restriction that an automatic system 
can never win back more than users have already taken. Without this restriction,
$|\pi_a|$ would be steeper.}
\label{rates}
\end{figure}
Thus, in summary, the characteristic matrix is given by:
\beq
\pi(u,s) = \left( 
\begin{array}{cccc}
-\2+\pi_g(t)  & -\2+\pi_g(t)        & -\2+\pi_g(t) & \pi_q\\
\2+\pi_u(t) & \2 + \pi_u(t) + \pi_a(t) & \2 +\pi_u(t) +\pi_a(t)\,\theta(t_0-t) & \pi_q\\
\2+\pi_u(t) & \2 + \pi_u(t)        & \2 +\pi_u(t)                      & \pi_q\\
\2+\pi_u(t) & \2 + \pi_u(t)        & \2 +\pi_u(t) +\pi_a(t)\,\theta(t_0-t) & \pi_q
\end{array}
\right),
\eeq
where the step function is defined by,
\beq
\theta(t_0-t) = \left\lbrace \begin{array}{c}
1 ~ (t \ge t_0)\\
0 ~ (t < t_0)
\end{array}
\right.,
\eeq
and represents the time-delay in starting the automatic tidying
system in the case of tidy-above-threshold. 

It is possible to make several remarks about the relative sizes of
these contributions. The automatic system works at least as fast as any
human so, by design, in this simple model we have
\beq
\2 \ge |\pi_a| \ge |\pi_u| \ge |\pi_g| \ge 0,
\eeq
for all times. In addition , for short times $\pi_q > \pi_u$, but
users can quickly fill their quota and overtake this.
In a zero-sum game, the automatic system can never tidy garbage
faster than users can create it, so the first inequality is
always saturated. From the nature of the cumulative pay-offs, we can 
also say that
\beq
(\2+\pi_u) \ge (\2 + \pi_u + \pi_a\theta(t_0-t))\ge (\2+\pi_u+\pi_a),
\eeq
and 
\beq
|\2+\pi_u| \ge |\pi_g - \2|.
\eeq
Let us now apply these results to a modest strategy of automatic tidying,
of garbage, once per day, in order to illustrate the utility of the game
formulation. The first step is to compute the pay-off rate contributions.
Referring to figure \ref{rates}, one sees that the automatic system can
always match users' moves. As drawn, the daily ripples of the automatic
system are in phase with the users' activity. This is not realistic, since
tidying would normally be done at night when user activity is low, however
such details need not concern us in this illustrative example. 

The policy we have created in setting up the rules of play for the
game, penalizes the system administrator for employing strict quota
shares.  Even so, users do not gain much from this, because quotas are
constant for all time.  A quota is a severe handicap to users in the
game, except for very short times before users reach their quota
limits. Quotas could be considered cheating in such a game, since they
determine the outcome even before play commences. There is no longer a
contest. Moreover, comparing the values in the figure, it is possible
to see how resource inefficient quotas are.  Users cannot create
temporary files which exceed these hard and fast quotas. An immunity
type model which allows fluctuations is a considerably more resource
efficient strategy, since it allows users to span all the available
resources for short periods of time, without consuming them for ever. 

Any two-person zero-sum game has a solution, either in terms of a pair
of optimal {\em pure} strategies or as a pair of optimal {\em mixed}
strategies\cite{morgenstern,dresher}. This result is known as the
{\em minimax} theorem and was proved by Von Neumann. The solution
is found as the balance between one player's attempt to maximize his
pay-off and the other player's attempting to minimize the opponent's result.

In general one can say of the pay-off matrix that
\beq
\max_{\downarrow}
\min_\rightarrow \pi_{rc}\le \min_\rightarrow \max_\downarrow \pi_{rc},
\eeq
where the arrows refer to the directions of increasing rows ($\downarrow$) 
and columns ($\rightarrow$).
The left hand side is the least users can hope to win (or conversely
the most that the system administrator can hope to keep) and the right
is the most users can hope to win (or conversely the least the system
admin can hope to keep).
If we have
\beq
\max_{\downarrow}\min_\rightarrow \pi_{rc} = \min_\rightarrow\max_\downarrow \pi_{rc},
\label{minimax}
\eeq
it implies the existence of a pair of single, pure strategies
$(r^*,c^*)$ which are optimal for both players, regardless of what the
other does.  If the equality is not satisfied, then the minimax
theorem tells us that there exist optimal mixtures of strategies,
where each player selects at random from a number of pure strategies
with a certain probability weight.

The situation for our time-dependent example matrix is different for
small $t$ and for large $t$. The distinction depends on whether users
have had time to exceed fixed quotas or not; thus `small $t$' refers
to times when users are not impeded by the imposition of quotas.

For small $t$, we have:
\beq
\max_{\downarrow}\min_\rightarrow \pi_{rc} &=& \max_{\downarrow}
\left(
\begin{array}{c}
\pi_g - \2\\
\2 + \pi_u + \pi_a\\
\2 + \pi_u\\
\2+\pi_u+\pi_a\,\theta(t_0-t)
\end{array}
\right)\nonumber\\
&=& \2+\pi_u.
\eeq
The ordering of sizes in the above minimum vector is:
\beq
\2+\pi_u \ge \2 + \pi_u +\pi_a\theta(t_0-t)\ge \pi_u +\pi_a\theta(t_0-t) \ge \pi_g-\2.
\eeq
This is useful to know, if we should examine what happens when
certain strategies are eliminated. For the opponent's endeavours
we have
\beq
\min_{\rightarrow}\max_\downarrow \pi_{rc} &=& \min_{\rightarrow}
(\2+\pi_u, \2+\pi_u, \2+\pi_u, \pi_q)\nonumber\\
&=& \2 + \pi_u.\label{w}
\eeq
This indicates that the equality in eqn. (\ref{minimax}) is satisfied
and there exists at least one pair of pure strategies which is optimal for both
players. In this case, the pair is for users to conceal files,
and for the system administrator to tidy by any means (these all
contribute the same weight in eqn (\ref{w}). Thus for small times,
the users are always winning the game if we assume that they are allowed
to bluff by concealment. If the possibility of concealment or bluffing
is removed (perhaps through an improved technology used by the administrator),
then the next best strategy is for users to bluff by changing the date.
In that case, the best system administrator strategy is to tidy at threshold.

These results make qualitative sense and tally well with the author's experience.
The result also makes a prediction for system administration tools like
cfengine. Cfengine must be able to see through attempts at bluffing if it is
to be an effective opponent against the worst users.

For large times (when system resources are becoming or have become scarce),
then the situation looks different. In this case one finds that
\beq
\max_{\downarrow}\min_\rightarrow \pi_{rc} = \min_\rightarrow\max_\downarrow \pi_{rc} = \pi_q.
\eeq
In other words, the quota solution determines the outcome of the game
for any user strategy. As already commented, this might be considered
cheating or poor use of resources, at the very least. If one eliminates
quotas from the game, then the results for small times hold also at
large times.

This simple example of system administration as a strategic game
between users and administrators was not intended to be as realistic
as possible, rather it was intended as an illustration of the
principles involved.  Nevertheless, it is already clear that user
bluffing and system quotas are strategies which are to be avoided
in an efficient system. By following this basic plan, it should be
possible to analyze more complex situations in future work.

\subsection{The policy $P(t)$ and the pay-off matrix $\pi(t)$}

At the beginning of this paper, we referred to a central axiom which
involved the changing system policy $P(t)$. The characteristic
(pay-off) matrix $\pi_{rc}(t)$ must clearly be related to this policy.

Let us suppose that the pay-off matrix is a $u\times s$ matrix, with
$u$ user strategies and $s$ system strategies.  The administrators
strategies are limited by the policy, and the rewards are also
limited, so both the dimension $s$ and form of the pay-off matrix
are functions of the policy. The user's strategies cannot be assumed
to be limited by policy however, since `criminal' users will ignore
policy for personal gain. Although one may think of the dimension
$s[P(t)]$ as being a functional of the policy, it would not be correct
to think of $u$ as being a functional of the policy, since there can
be no restriction on what users will try, simply as a result of law-giving.
User's actions can only be restricted by applying counter-measures
within the
\beq
\pi= \pi_{rs[P(t)]}(P(t)).
\eeq
It should be noted, however, that there can be no unique mapping between
policy and pay-off matrix.

\subsection{Change and future models}

Expressing deterministic changes in generic computer systems would be
a huge undertaking unless one restricted ambitions to general features
and trends. Dynamical systems are difficult to trace, even in the
simplest of cases, so one cannot expect to get very far without making
significant simplifications. The aim of considering a dynamical theory
is thus to characterize the significant trends of change which might
occur, owing to idealized influences. A full discussion of this topic
is beyond the scope of the present paper, however based on the axioms
and deliberations presented here, it is possible to outline the way 
forward in studying them.

The expression of strategies in the previous section is too general to
be useful for a fully general, dynamical theory. Taking account of
every strategic detail would be a vast undertaking. Instead, one can
analyze the development at the level of a generic computer system
undergoing generic changes as a matter of principle.  The purpose of
such a vague preliminary investigation is to elucidate the
relationship between the system administration game and the lattice
description of the ideal state, presented in section 8.

Once a strategy mixture has been decided, one must address the fact
that, in real-world games, the speed of information is finite. It will
take a finite amount of time for a response to develop after a
strategy is implemented. Moves and counter-moves do not follow a rigid
time-plan as in games like chess. This kind of delay leads to races and
duels for superiority between competing players. Delay is the province
of linear response theory.

The aim, then, is to express the causal structure of system development 
in the foregoing mathematical language.  In order to reduce the
dynamical game to algebra we must express each of these in terms of
basic primitives.  Causality is about relating actions to outcomes, or
changes of state $\delta S$.  A general action $A(t)$ is built up from a
number of primitive action-types $T_a$ (called the generators for the
action transformation) in a linear combination
\beq
A(t) = \sum_i\; a_i(t)\,T_i
\eeq
where $a_i(t)$ are functions of time (not necessarily differentiable,
often step-like) and $i$ takes values which number the
full spectrum of primitive actions.  The $T^i$ are orthogonal vectors
or matrices (indices suppressed), one for each primitive action type,
which span an abstract vector space. This vector space is the chequerboard
on which the game takes place.

Each complete action $A(t)$, results in a change in the state of the
system, which may be denoted $\delta S$.  An action can also be a
causal chain of sub-actions, characterizing a sequence of changes in
the state. This type of causal relationship is summarized by a {\em
Green function}, {\em propagator}, or {\em response-function}
formulation\cite{feynman1,schwinger1}:
\beq
\delta S(t) = \int dt' G(t,t')\,A(t'),
\eeq
where $G(t,t')$ is the two-point response function, as yet unspecified.
If the rules of the game are independent of time, then $G(t,t')=G(t-t')$;
if the rules change over time, then $G(t,t')=G(t-t',t+t')$.
In this language of dynamical systems, an action plays the role
of a {\em source} or {\em driving force} for the system.  The equation
above may be inverted to provide an inhomogeneous differential equation for
the changing state of the system. If one formally introduces a 
differential operator ${D}_t$ which is the inverse of the response function:
\beq
\int dt'\; D_t\; G(t,t') = 1,
\eeq
then the differential equation may be written, schematically:
\beq
D_t S(t) = A(t),
\eeq
where, as {\em ad hoc} an example one might have,
\beq
D_t \equiv \frac{d^2}{dt^2} + i\gamma \frac{d}{dt} + \omega_0^2,
\eeq
for an approximately periodic system which degrades over time, like
a damped harmonic oscillator.
Each action $A(t)$ thus leads to a response or change of state; this
in turn implies that the state of the system must be a linear
combination of the same action types:
\beq
S(t) = \sum_i s_i(t)\, T_i.
\eeq
The state is thus defined on the same lattice, or chequerboard as the
actions themselves. Differential (difference) characterizations of
state have been studied in ref. \cite{huberman1}; this type of
description is interesting, since it leads often to rich
dynamics. Alternating periods of change and stability (riffles and
pools in the flow of the system) might be best described by a
difference representation.

Returning to the idea of the contest as a game, one writes a strategy
as a statistical mixture of actions (i.e. moves in the game) $A(t)$,
applied over an interval of time. This stochastic mixture specifies
the boundary conditions under which the actions are applied.  It may
be formed as a linear combination of basic actions $A_n$, with
probability weights $w_n$:
\beq
J(t) = \sum_n w_n A_n(t) = \sum_i p_i T_i,
\eeq
The strategy vector $J_i$ is the vector of probabilities for each
primitive action, given the chosen mixture of full actions $A_n$
for $J$. In other words, $J_i$ are the components of the decomposition
of the strategy $J(t)$ on the space of primitive actions.
\beq
p_i = \sum_n w_n.
\eeq
It is easy to normalize these so as to be actual probabilities which sum to
unity
\beq
\sum_i p_i \equiv 1.
\eeq
Notice that the specific representation of basis generators $T_i$ does not
affect the strategy vector, since it only serves to label the
lattice-work of independent actions. There is no 
unique labelling.  The components with respect to the basis must
be related by a response function $\Pi_{ij}(t,t')$
\beq
\delta S_i = \int dt' \; \Pi_{ij}(t,t')J_j(t').
\eeq
The matrix value distribution is related to the pay-off matrix, and a
basis of so-called ladder operators, also called creation and
annihilation operator. The represent do-action/undo-action operations
of the users and system administrator.
\beq
\Pi(t,t') \sim \pi_{ij}\; \otimes\; \langle \vec d |\hat S_+(t) \hat S_-(t') |\vec d \rangle
\eeq
where $\hat S_\pm$ are operators which annihilate a configuration
state at $t'$ and create a new configuration at time $t$. This is the
generic mechanism by which the system develops. This form of
description might seem unnecessarily formal, but it is actually highly
useful, since the continuous generalization of this kind of dynamical
system has been widely studied in statistical field theory. By picking
out universal features of statistical models and restricting the scope
of the computer system, there is a real chance of being able to build
toy models which have qualitative, predictive power. However, this is
no trivial undertaking and will be considered in a later paper\cite{inpreparation}.

Since the actions which configure a computer system form a lattice,
and these primitive action types do not necessarily commute with one
another, one concludes that a suitable idealization of the system
administration's stochastic dynamics is found in non-Abelian, statistical field
theories. This line of study would be suitable for modelling resource
availabilities for large numbers of users, in which all users behave
approximately equally on average (like an ideal gas).  This approach
promises therefore to be relevant to the problem of anomaly
detection\cite{burgessPRL} and will be returned to in later work.

\section{Summary}

The aim of this paper has been to formulate a trustworthy framework
for analyzing models of system administration. There is good cause to
view computers as dynamical systems, approximated by mechanistic rules
developing in time, with idealized properties which can be summarized
by a finite state lattice. The theory of games has been employed in
order to select between alternative strategies in a contest for
machine resources, moving the state of the system through the lattice,
as if on a chequerboard.  It has been shown that it is possible to
see system administration as the effort to keep the system close to an
ideal state, by introducing countermeasures in the face of competitive
resource consumption. This is the formal basis which opens the way for
objective analyses in the field.

It is important to understand that, even an answer obtained with the
assistance of a mathematical formalism is not necessarily the last
word on the subject.  Mathematics is only a tools for relating
assumptions to conclusions, in an impartial way. With a mathematical
approach, it becomes easier to see through personal opinions and
vested interests when assumptions and methods are clearly and
rigorously appraised.  However, one can only distinguish between those
possibilities which are taken into account. That means that every
relevant strategy, or alternative, has to be considered, or else one
could miss the crucial combination which wins the game. This is the
limitation of game theory.  It is not generally possible to determine
strategies without creative input; this means that human intelligence
will be required for the foreseeable future. There can be no
zero-maintenance computer system.  With this caution, how can one know
that the ideal state of a system can be reached? How can one know that
the system will not run away in an unstable spiral to catastrophe?

Two things are clear from the limited analysis here. The first is that
purely dumb automatic systems are inadequate to perform every task in
system administration today. Intelligent incursions are required to
solve complex problems, to extend or adjust the strategies of the
automatic system.  Interestingly, this is the approach by which
evolution has solved the immunity problem: the automatic responses of
lymphocytes only go so far; the emergence of intelligence in humans
has enabled us to develop medical research and develop drugs and other
treatments against damage and disease.  It seems naive to believe that
any simple mechanistic system would be able to do any better than
this; we can expect to require the assistance of humans at least until
alternative machine intelligences have been developed.

The second point is that the use of quotas is a highly inefficient way
of counteracting the effects of selfish users. A quota strategy can
never approach the same level of productivity as one which is based on
competitive counterforce. The optimal strategies for garbage
collection are rather found to lie in the realm of the immunity
model\cite{burgessLISA99}.  However, it is a sobering thought that a
persistent user, who is able to bluff the immune system into
disregarding it, (like a cancer) will always win against the resource
battle. The need for new technologies which can see through bluffs
will be an ever present reality in the future. With the ability of
encryption and compression systems to obscure file contents, this
is a contest which will not be easily won by system administrators.

There is plenty of work to be done on the theory of system administration.
This paper is merely a small push in the direction of progress.

I am grateful to Trond Reitan for a useful discussion about evolutionary
stable strategies and to H\aa rek Haugerud, Lars Kristiansen and 
particularly Sigmund Straumsnes
for their critical readings of the manuscript.


\begin{thebibliography}{10}

\bibitem{evard1}
R.~Evard.
\newblock An analysis of unix system configuration.
\newblock {\em Proceedings of the 11th Systems Administration conference
  (LISA)}, page 179, 1997.

\bibitem{burgessLISA99}
M.~Burgess, H.~Haugerud, and S.~Straumsnes.
\newblock Measuring host normality.
\newblock {\em Software Practice and Experience (submitted)}, 1999.

\bibitem{burgessLISA98}
M.~Burgess.
\newblock Computer immunology.
\newblock {\em Proceedings of the 12th Systems Administration conference
  (LISA)}, page 283, 1998.

\bibitem{burgessbook}
M.~Burgess.
\newblock {\em Principles of Network and System Administration}.
\newblock J. Wiley \& Sons, Chichester, 2000.

\bibitem{NASAsteve}
S.~Traugott and J.~Huddleston.
\newblock Bootstrapping an infrastructure.
\newblock {\em Proceedings of the 12th Systems Administration conference
  (LISA)}, page 181, 1998.

\bibitem{burgess1}
M.~Burgess.
\newblock A site configuration engine.
\newblock {\em Computing systems}, 8:309, 1995.

\bibitem{huberman1}
N.~Glance, T.~Hogg, and B.A. Huberman.
\newblock Computational ecosystems in a changing environment.
\newblock {\em International Journal of Modern Physics}, C{\bf 2}:735, 1991.

\bibitem{burgessPRL}
M.~Burgess.
\newblock Thermal, non-equilibrium phase space for networked computers.
\newblock {\em Physical Review E (submitted)}, 1999.

\bibitem{burgess3}
M.~Burgess and D.~Skipitaris.
\newblock Adaptive locks for frequently scheduled tasks with unpredictable
  runtimes.
\newblock {\em Proceedings of the 11th Systems Administration conference
  (LISA)}, page 113, 1997.

\bibitem{morgenstern}
J.V. Neumann and O.~Morgenstern.
\newblock {\em Theory of games and economic behaviour}.
\newblock Princeton University Press, Princeton, 1944.

\bibitem{dresher}
M.~Dresher.
\newblock {\em The mathematics of games of strategy}.
\newblock Dover, New York, 1961.

\bibitem{nash1}
J.F. Nash.
\newblock {\em Essays on Game Theory}.
\newblock Edward Elgar, Cheltenham, 1996.

\bibitem{maynardsmith}
J.~Maynard-Smith.
\newblock {\em Evolution and the Theory of Games}.
\newblock Cambridge University Press, Cambridge, 1981.

\bibitem{lisa8941}
E.D. Zwicky.
\newblock Disk space management without quotas.
\newblock {\em Proceedings of the third systems administration conference LISA,
  (SAGE/USENIX)}, page~41, 1989.

\bibitem{feynman1}
R.P. Feynman.
\newblock Spacetime approach to quantum electrodynamics.
\newblock {\em Reviews of Modern Physics}, {\bf 20}:267, 1948.

\bibitem{schwinger1}
J.~Schwinger.
\newblock The theory of quantized fields 1.
\newblock {\em Physical Review}, {\bf 82}:914, 1951.

\bibitem{inpreparation}
M.~Burgess.
\newblock In preparation.

\end{thebibliography}
\end{document}